\documentclass[10pt,conference]{IEEEtran}
\usepackage{amsmath,amssymb,amsfonts}
\usepackage{algorithmic}
\usepackage{listings}
\usepackage{graphicx}
\usepackage{textcomp}
\usepackage{biblatex}
\def\BibTeX{{\rm B\kern-.05em{\sc i\kern-.025em b}\kern-.08em
    T\kern-.1667em\lower.7ex\hbox{E}\kern-.125emX}}

\lstdefinelanguage{OpenQASM}
    {morekeywords={gate,OPENQASM,qubit,reset,measure,h,U,cx,CX,bit,if,else},
    morecomment=[l]{//},
}

\lstdefinelanguage{MLIR}
    {morekeywords={module,quir,oq3,qcs,pulse,scf,return,func,arith},
}

\makeatletter
\def\lst@makecaption{%
  \def\@captype{table}%
  \@makecaption
}
\makeatother

\makeatletter 
\newcommand{\linebreakand}{%
  \end{@IEEEauthorhalign}
  \hfill\mbox{}\par
  \mbox{}\hfill\begin{@IEEEauthorhalign}
}
\makeatother 

\begin{document}

\title{Design and architecture of the IBM Quantum Engine Compiler}

\author{\IEEEauthorblockN{Michael B. Healy\IEEEauthorrefmark{1},
Reza Jokar\IEEEauthorrefmark{1},
Soolu Thomas\IEEEauthorrefmark{1},
Vincent R. Pascuzzi\IEEEauthorrefmark{1},
Kit Barton\IEEEauthorrefmark{2},\\ 
Thomas A. Alexander\IEEEauthorrefmark{2},
Roy Elkabetz\IEEEauthorrefmark{3},
Brian C. Donovan\IEEEauthorrefmark{4},
Hiroshi Horii\IEEEauthorrefmark{5} and
Marius Hillenbrand\IEEEauthorrefmark{6}}
\IEEEauthorblockA{\IEEEauthorrefmark{1}\textit{IBM Quantum},
\textit{T.J. Watson Research Center},
Yorktown Heights, NY, USA\\
mbhealy@ibm.com, jokar@ibm.com, soolu.thomas@ibm.com, vrpascuzzi@ibm.com}
\IEEEauthorblockA{\IEEEauthorrefmark{2}\textit{IBM Quantum}, 
\textit{IBM Canada Software Lab}, 
Markham, ON Canada\\
kbarton@ca.ibm.com, thomas.a.alexander@ibm.com}
\IEEEauthorblockA{\IEEEauthorrefmark{3}\textit{IBM Quantum},
\textit{IBM Research Israel},
Haifa, Israel \\
roy.elkabetz@ibm.com}
\IEEEauthorblockA{\IEEEauthorrefmark{4}\textit{IBM Quantum},
\textit{IBM Research Cambridge},
Cambridge, MA USA \\
brian.donovan@ibm.com}
\IEEEauthorblockA{\IEEEauthorrefmark{5}\textit{IBM Quantum},
\textit{IBM Research Tokyo},
Tokyo, Japan\\
horii@jp.ibm.com}
\IEEEauthorblockA{\IEEEauthorrefmark{6}\textit{IBM Quantum},
\textit{IBM B\"{o}blingen Laboratory},
B\"{o}blingen, Germany\\
marius.hillenbrand@ibm.com}
}

\maketitle

\begin{abstract}
In this work, we describe the design and architecture of the open-source Quantum Engine Compiler (\textit{qe-compiler}) currently used in production for IBM Quantum systems. The \textit{qe-compiler} is built using LLVM's Multi-Level Intermediate Representation (MLIR) framework and includes definitions for several dialects to represent parameterized quantum computation at multiple levels of abstraction. The compiler also provides Python bindings and a diagnostic system. An open-source LALR lexer and parser built using Bison and Flex generates an Abstract Syntax Tree that is translated to a high-level MLIR dialect. An extensible hierarchical target system for modeling the heterogeneous nature of control systems at compilation time is included. Target-based and generic compilation passes are added using a pipeline interface to translate the input down to low-level intermediate representations (including LLVM IR) and can take advantage of LLVM backends and tooling to generate machine executable binaries. The \textit{qe-compiler} is built to be extensible, maintainable, performant, and scalable to support the future of quantum computing.
\end{abstract}


\section{Introduction} \label{sec:intro}
Control systems for quantum computers must continue to scale up to support the large number of qubits needed for error correction and to enable future applications. Software systems designed to support large numbers of qubits will also need to support a large system of control electronics. Many approaches to quantum control electronics utilize networked systems of controllers implemented using classical computing hardware driven by customized real-time processor architectures. These processors in turn must execute instruction streams that implement a user's specified algorithm. Generating these instruction streams is the goal of a low-level quantum control system compiler, often referred to as a backend compiler.

In this work, we introduce IBM's open-source Quantum Engine Compiler (\textit{qe-compiler}) \cite{qe-compiler}, which forms the core of the backend compiler used in IBM Quantum's production Quantum Service. The \textit{qe-compiler} supports dynamic circuit execution described using OpenQASM 3~\cite{10.1145/3505636} input and is based on the LLVM Multi-Level Intermediate Representation (MLIR) framework \cite{mlir}. The compiler is mainly written in C++17 and includes a Python interface for generating target-independent payloads as well as MLIR dialect-level Python interfaces for generating MLIR that can be passed directly to the core pass pipelines. Additional interfaces are provided for supporting the parametric~\cite{Cerezo_2021} compilation of circuits. An existing payload can be \textit{linked} with a new set of parameter values to generate rapid program updates without the need for recompilation. The compiler also contains an extensible hierarchical target system model supporting the definition of target-based passes. A threaded compilation manager is provided that enables parallelized compilation of target-specific pass pipelines across the target system, which enables scaling to large systems.

The \textit{qe-compiler} has been designed and built by developers working closely with the wider quantum engine service team. This includes the hardware groups responsible for the architecture of the quantum control systems, the execution team, and the runtime team. The compiler has been designed and used mainly in the context of IBM Quantum's transmon-based systems, but is general and flexible enough to be useful in the control system stacks of other types of quantum systems.

\begin{figure}[!t]
\centering
\includegraphics[width=75mm]{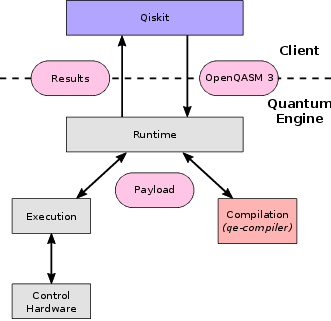}
\caption{The architecture of the quantum engine. The compiler accepts OpenQASM 3 input and returns an executable payload to the runtime service.}
\label{fig:quantum-engine-architecture}
\end{figure}

Figure~\ref{fig:quantum-engine-architecture} shows a diagram of the compiler's position within the wider quantum engine software stack. Users send a job to the service using OpenQASM 3, the runtime then calls the compiler with the user's input, which produces an executable payload. The runtime provides the compiled payload to the execution service, which uses drivers to load the payload into the control system and then execute it. The drivers return result data to the execution service. The execution service passes the results back to the runtime service and on to the user.

The rest of this document is structured as follows. In Section~\ref{sec:target-system} we discuss the details of the target system. Section~\ref{sec:parser} describes the lexer and parser. Section~\ref{sec:IRs} details the custom IRs the compiler is built upon and includes an example OpenQASM 3 program converted into MLIR. Section~\ref{sec:lowering} presents an example demonstrating the lowering pipeline using a mock target and describes the threaded compilation manager. Section~\ref{sec:payload} describes the payload framework. Section~\ref{sec:python-interface} describes the various Python interfaces for the compiler, and Section~\ref{sec:diagnostics} discusses the diagnostic system. Finally, we provide concluding remarks in Section~\ref{sec:conclusion} and list acknowledgements in Section~\ref{sec:ack}.

\section{Target System} \label{sec:target-system}

\begin{figure}[!t]
\centering
\includegraphics[width=70mm]{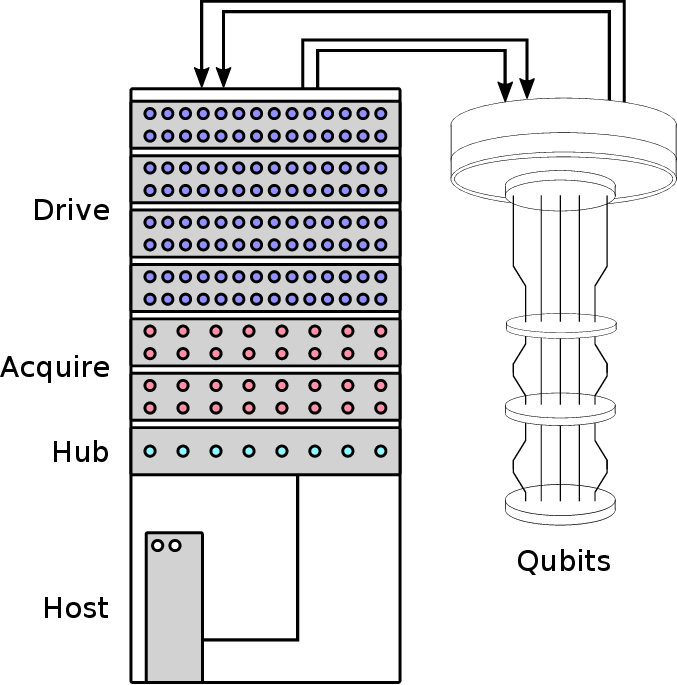}
\caption{Quantum control system structure. Control electronics on the left drive and receive signals from the dilution refrigerator on the right holding the qubits.}
\label{fig:quantum-control-system}
\end{figure}

Control systems for superconducting qubits are built from waveform generators and receivers. These drive and acquire controllers are connected together by a hub that can perform classical computation on received measurement data and send control flow information back to the controllers. The host loads the instruments with each job, triggers the start of execution, monitors progress, and collects result data to return to users after execution completes. The backend compiler is responsible for producing the instruction streams that implement each job requested by a user on each instrument in the system. Figure~\ref{fig:quantum-control-system} shows a diagram of a control system with a rack of control electronics on the left, and the dilution refrigerator holding the qubits on the right.

The compiler models a quantum control system as a \textit{target system} that forms a hierarchical description of the control system as a tree graph. Each object in the tree is based on a top-level parent \texttt{Target} class. This contains generic information that all targets must have such as a name. The \texttt{Target} class also owns a set of \textit{child targets}, which may be accessed when traversing the target system tree. Lastly, there are methods for adding MLIR module content to a payload (Section \ref{sec:payload}), as well as methods for adding and returning diagnostics safely with threading (Section \ref{sec:diagnostics}).

\begin{figure}[!t]
\centering
\includegraphics[width=83mm]{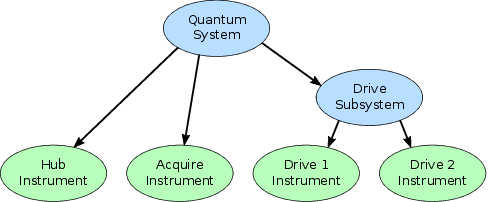}
\caption{Example target system hierarchy for a quantum system with two drive instruments, an acquire instrument, and a hub instrument.}
\label{fig:TargetSystemEx}
\end{figure}

The \texttt{Target} class is then specialized with two derived classes, the \texttt{TargetSystem} and the \texttt{TargetInstrument}. \texttt{TargetSystems} are intended to represent higher-level collections of instruments and subsystems with customized orchestration logic across them (\textit{e.g.}, during system initialization or instrument synchronization). For example, a \texttt{DriveSubsystem} could be a child of a top-level control system architecture while also being the parent of all drive instruments within that control system. \texttt{TargetInstruments} are intended to be leaf nodes of the target tree graph, representing individual controllers requiring content within a payload. Figure \ref{fig:TargetSystemEx} shows an example hierarchy for a control system that drives two qubits and acquires data from those two qubits using a third instrument.

\section{OpenQASM 3 Lexer and Parser} \label{sec:parser}

The \textit{qe-compiler} uses a companion open-source project \textit{qe-qasm}~\cite{qe-qasm} to perform lexical analysis and parse input OpenQASM 3 programs. This project is built on the Flex and Bison lexer and parser generators to create a LALR(1) parser and abstract syntax tree (AST) generator. The \textit{qe-compiler} then walks the AST to generate MLIR corresponding to the input program. The \textit{qe-qasm} parser currently supports most programs written in OpenQASM 2.0, as well as the low-level features of the OpenQASM 3 specification, and was carefully written to avoid shift-reduce and reduce-reduce conflicts that may cause ambiguities. The AST generator uses the GNU MPFR~\cite{mpfr} and MPC~\cite{mpc} libraries to support arbitrary precision floating point and complex literals. 

The \textit{qe-qasm} parser also contains a robust diagnostic system with support for setting handlers in calling code. This enables the \textit{qe-compiler} to connect its diagnostic system and forward diagnostic warnings and errors up the quantum software stack (see Sections \ref{sec:python-interface} and \ref{sec:diagnostics}).

\section{Specialized Intermediate Representations} \label {sec:IRs}

The MLIR project within LLVM enables a flexible definition of an arbitrary number of IRs, called dialects, along with automated support for verification, printing, parsing, pass definitions and management, and transformations. It is the basis for several existing quantum IRs \cite{9605269, pennylane-catalyst, 10.1145/3497776.3517772, The_CUDA_Quantum_development_team_CUDA_Quantum} and forms the basis of the \textit{qe-compiler}. MLIR dialects, like most of the LLVM project, uses static single assignment (SSA) form, which eases the creation of many optimizations and analyses. MLIR is structured using \textit{operations} that take \textit{arguments} and produce \textit{values}. Each operation can recursively contain zero or more \textit{regions} containing \textit{blocks} that hold additional operations. MLIR includes many built-in dialects for classical computing, along with transformations between them and into LLVM IR. The \textit{qe-compiler} contains definitions for several dialects employed at different stages of compilation, the OpenQASM 3 (OQ3) IR, the Quantum Control System (QCS) IR, the Quantum IR (QUIR), and the Pulse IR. The following subsections describe each in more detail.

\subsection{OpenQASM3 IR} \label{sec:oq3IR}

The OpenQASM 3 (\texttt{oq3}) dialect is targeted at representing the high-level types and operations special to the OpenQASM 3 language~\cite{10.1145/3505636}. These include operations creating the special types defined by the language like the classical bit (cbit), angles, stretches, and arrays, as well as mathematical and logical operations upon them. Casting between types is accomplished using the \texttt{cast} operation (\textit{e.g.} conversions between classical bits and integers). The OQ3 dialect provides many operations that interact with the QUIR dialect types (Section~\ref{sec:QUIR}). We have also defined several operations for general variable handling, which help with the conversion from OpenQASM 3 semantics into the SSA form required by the lower-level dialects.

The variable handling operations are:
\begin{itemize}
    \item \texttt{declare\_variable}, creates new variables with a given symbol name,
    \item \texttt{variable\_assign}, assigns a value to an existing variable, and
    \item \texttt{variable\_load}, returns the current value of an existing variable.
\end{itemize}
Together, these operations define in MLIR the semantics of a C-like language where all variables represent objects or types in memory. Transformations are defined within the compiler that convert these operations into equivalent built-in MLIR operations and types, such as the \texttt{memref} dialect's \texttt{global} operation. This enables the compiler to take advantage of normal global memory allocation and use optimizations for OQ3 types. This system currently does not handle the variable scoping rules of OpenQASM 3 exactly as all variable names are global. Future updates will focus on improving this support, potentially using scope-mangled variable names.

The compiler currently supports only fully transpiled and mapped quantum circuits. Only non-virtual, purely physical qubits (those defined by the \texttt{\$} symbol in OpenQASM 3) are supported as input.

\subsection{Quantum Control System IR} \label{sec:qcsIR}

The Quantum Control System (\texttt{qcs}) dialect defines operations for important functions of quantum control systems. These include operations for initializing and finalizing the control system and shot loops, synchronizing the elements of the control system, broadcasting, sending, and receiving values, handling job-level parameters, and representing parallel control flow within the control system. 

In particular, the \texttt{parallel\_control\_flow} operation holds a single-block region wherein each operation is a control flow operation (from the MLIR structured control flow (\texttt{scf}) dialect) that will execute in parallel in different components of the control system. This is used primarily for parallelizing reset operations but can in principle be used to represent any form of non-uniform parallelizable control flow.

OpenQASM 3 supports both \textit{input} and \textit{output} keywords for declaring that a classical value is an input or output of a program. The QCS dialect supports these declarations using the \texttt{declare\_parameter} operation, and the current value is accessed using the \texttt{parameter\_load} operation.

\subsection{Quantum IR} \label{sec:QUIR}

The Quantum IR/dialect (\texttt{quir}) represents quantum circuits as applied to qubits with memory semantics. It defines special types, such as classical bits, quantum bits, angles, durations, and stretches, though many of the defining operations are currently contained in the OQ3 dialect (Section \ref{sec:oq3IR}) to which these will eventually be migrated. The built-in qubit operations are represented by specific operations (\texttt{barrier}, \texttt{builtin\_CX}, \texttt{builtin\_U},  \texttt{delay}, \texttt{measure}, \texttt{reset}). We also include an operation for calls to user-specified gates (\texttt{call\_gate}) and an operation for obtaining constants of the special QUIR types (\texttt{constant}). Finally, we use a special operation (\texttt{declare\_qubit}) for declaring the existence of a qubit with a specific ID that produces a qubit value that can be used in the rest of the circuit in a memory-like model. Custom gate definitions are modeled using normal MLIR function operations. 

OpenQASM 3 programs are lexicographically ordered quantum/classical operations interspersed with control flow operations which themselves may contain blocks of quantum circuit operations and classical computations. We explicitly separate the quantum operations from control flow and classical computations to enable code reuse through circuit invocation, which simplifies many of the passes within the compiler, including scheduling. To do this, we have created operations for circuit definition (\texttt{circuit}) and circuit invocation (\texttt{call\_circuit}). After initial MLIR generation, a pass is applied that gathers all quantum operations into \texttt{circuit}s to create the canonical form for later passes. While \texttt{circuit}s are treated specially in the compiler by many passes, their function and definition are nearly identical to normal functions and make use of MLIR's function interfaces.

\subsection{Pulse IR} \label{sec:PulseIR}

The Pulse (\texttt{pulse}) dialect is targeted at representing the lowest level of quantum operations within the compiler consistent with the OpenPulse specification~\cite{mckay2018qiskit}. Quantum operations (gates and measurement) are translated into pulse-level operations before pulse scheduling and final lowering to hardware dialects. At the pulse level, we convert quantum circuits into sequences of pulse operations on frames using MLIR pulse calibrations that the compiler receives as input. The analogue to QUIR's \texttt{circuit} and \texttt{call\_circuit} are the Pulse's \texttt{sequence} and \texttt{call\_sequence}, respectively, which hold (or invoke) statically schedulable sequences of pulse operations that implement quantum circuits.

In the \texttt{pulse} dialect \textit{frames} are abstractions that act as both a clock within the quantum program, with time being incremented on each usage, and a stateful carrier signal defined by a frequency and phase. \textit{Ports} are representations of physical channels provided by hardware vendors to manipulate and observe qubits. Frames can be mixed with ports to create mixed frames using the \texttt{mix\_frame} operation. We have defined analogs of many \texttt{quir} operations that act on mixed frames instead of qubits (\textit{e.g.}, \texttt{barrier} and \texttt{delay}). \texttt{shift\_phase} and \texttt{set\_phase} shift and set the phase of mixed frames, respectively, and \texttt{shift\_frequency} and \texttt{set\_frequency} shift and set the frequency of mixed frames, respectively. The \texttt{play} operation plays waveforms (created by \texttt{create\_waveform} operations) on mixed frames; each pulse being played has a waveform to define an envelop and a frame to track the frequency and phase. Measurements are converted into readout \texttt{play} operations followed by \texttt{capture} operations.

The compiler can accept a pulse MLIR program directly as input. Optimizing performance at the pulse level increases the fidelity of experiments, ultimately advancing the capabilities of quantum computers. 

\subsection{OpenQASM 3 to MLIR Example}

A simple example OpenQASM 3 program is shown below in Listing \ref{lst1:qasm3-example}. Three qubits are declared and then reset, followed by a Hadamard gate and measurement of hardware qubit \texttt{\$2}. This measurement value is then used to conditionally prepare qubits \texttt{\$0} and \texttt{\$1} in either a bell state or a non-entangled state before both qubits are measured at the end.

\begin{lstlisting}[
  caption=OpenQASM 3 example input,
  language=OpenQasm,
  label=lst1:qasm3-example,
]
OPENQASM 3.0;

gate cx c, t { CX c, t; }
gate h q {
  U(1.57079632679, 0.0, 3.14159265359) q;
}

// For now these declarations are required
qubit $0; qubit $1; qubit $2;

reset $0; reset $1; reset $2;

h $2;
bit mid = measure $2;
if (mid) {
  h $0;
  cx $0, $1;
} else {
  h $0;
  h $1;
}

bit[2] fin;
fin[0] = measure $0;
fin[1] = measure $1;
\end{lstlisting}

The program from Listing \ref{lst1:qasm3-example} run through the initial stages of the compiler (parsing and AST generation, followed by conversion to MLIR) and then canonicalized results in the MLIR program shown in Listing \ref{lst2:mlir-example}.


\begin{lstlisting}[
  caption=MLIR example output after walking the AST,
%  language=MLIR,
  label=lst2:mlir-example,
]
module {
  oq3.declare_variable @mid : !quir.cbit<1>
  oq3.declare_variable @fin : !quir.cbit<2>
  func.func @cx(%arg0: !quir.qubit<1>, 
                %arg1: !quir.qubit<1>) {
    quir.builtin_CX %arg0, %arg1 : 
                !quir.qubit<1>, !quir.qubit<1>
    return
  }
  func.func @h(%arg0: !quir.qubit<1>) {
    %angle = quir.constant #quir.angle<1.57079632679> : 
                !quir.angle<64>
    %angle_0 = quir.constant #quir.angle<0.000000e+00> : 
                !quir.angle<64>
    %angle_1 = quir.constant 
            #quir.angle<3.14159265359> : !quir.angle<64>
    quir.builtin_U %arg0, %angle, %angle_0, %angle_1 : 
                !quir.qubit<1>, !quir.angle<64>, 
                !quir.angle<64>, !quir.angle<64>
    return
  }
  func.func @main() -> i32 {
    %c0_i32 = arith.constant 0 : i32
    %c0_i2 = arith.constant 0 : i2
    %dur = quir.constant #quir.duration<1.000000e+00> : 
                !quir.duration<ms>
    %c1 = arith.constant 1 : index
    %c1000 = arith.constant 1000 : index
    %c0 = arith.constant 0 : index
    qcs.init
    scf.for %arg0 = %c0 to %c1000 step %c1 {
      quir.delay %dur, () : !quir.duration<ms>, () -> ()
      qcs.shot_init {qcs.num_shots = 1000 : i32}
      %0 = quir.declare_qubit {id = 0 : i32} : 
                !quir.qubit<1>
      %1 = quir.declare_qubit {id = 1 : i32} : 
                !quir.qubit<1>
      %2 = quir.declare_qubit {id = 2 : i32} : 
                !quir.qubit<1>
      quir.reset %0 : !quir.qubit<1>
      quir.reset %1 : !quir.qubit<1>
      quir.reset %2 : !quir.qubit<1>
      quir.call_gate @h(%2) : (!quir.qubit<1>) -> ()
      %3 = quir.measure(%2) : (!quir.qubit<1>) -> i1
      %4 = "oq3.cast"(%3) : (i1) -> !quir.cbit<1>
      oq3.variable_assign @mid : !quir.cbit<1> = %4
      %5 = oq3.variable_load @mid : !quir.cbit<1>
      %6 = "oq3.cast"(%5) : (!quir.cbit<1>) -> i1
      scf.if %6 {
        quir.call_gate @h(%0) : (!quir.qubit<1>) -> ()
        quir.builtin_CX %0, %1 : 
                    !quir.qubit<1>, !quir.qubit<1>
      } else {
        quir.call_gate @h(%0) : (!quir.qubit<1>) -> ()
        quir.call_gate @h(%1) : (!quir.qubit<1>) -> ()
      }
      %7 = "oq3.cast"(%c0_i2) : (i2) -> !quir.cbit<2>
      oq3.variable_assign @fin : !quir.cbit<2> = %7
      %8 = quir.measure(%0) : (!quir.qubit<1>) -> i1
      oq3.cbit_assign_bit @fin<2> [0] : i1 = %8
      %9 = quir.measure(%1) : (!quir.qubit<1>) -> i1
      oq3.cbit_assign_bit @fin<2> [1] : i1 = %9
    } {qcs.shot_loop}
    qcs.finalize
    return %c0_i32 : i32
  }
}
\end{lstlisting}

\section{Lowering to Hardware} \label{sec:lowering}

The compiler contains an example \texttt{TargetSystem} named the \texttt{MockTarget}. The Mock target includes a pipeline for lowering to the native target of the host system running the compiler. The generated binaries are non-functional and for demonstration only, but they provide an example for creating targets and integrating them into the compilation system. Our target pipelines directly produce binaries for each controller in the system. This is accomplished through many transformations of the input using the pass management system in MLIR and the threaded compilation manager that we have created to support parallelism and improve compilation speed. 

An important function within the compiler is to determine which parts of the input program are necessary to execute within each portion of the control system. While most program representations, such as OpenQASM 3, are in the form of a single input source, the actual programs are expected to be executed in a highly concurrent fashion across many controllers. It is necessary to break the input program apart and localize the portion of the source program to its respective \texttt{Target}. For example, classical computations should generally be executed on a portion of the control system with traditional computer architectures, gates should be executed on drive controllers, and, captures should be executed on acquire controllers, \textit{etc.}. Thus, compilation is generally divided into two phases. In the first phase, the overall structure of the input program is analyzed and optimized at a \texttt{TargetSystem} level. This is the phase where we combine statically schedulable gate sequences into the \textit{circuit}s described in Section~\ref{sec:QUIR}, as well as performing scheduling of those gate sequences within the global context of the program across the control system. After this phase, we break the input into units of code (called \textit{modules} in MLIR) that will be lowered to the target for each instrument in the system. We call this process \textit{localization}. After localization, there is one module for each instrument in the system containing the IR necessary to produce a payload for that instrument. These modules are completely independent from one another at this stage and may be processed in parallel. The post-localization phase of compilation is focused on lowering the IR to the point that it can be handed off to code generation pipelines.

Traditional compilers are focused on producing binaries for a single target architecture in each execution, this means that every pass has the potential to be useful in producing the final output. However, in the heterogeneous control systems used to drive quantum computers, it is often the case that entirely different pipelines are needed to lower the program IR to the final target architecture for that portion of the system. Traditional pass managers apply each pass in sequence to the input. Passes are often set up with an early check that decides if the pass should be applied given the target under consideration. The pass can then exit early if it will not generate any benefit. If multiple targets of the same architecture are desired, all passes will either execute or exit early simultaneously. However, with heterogeneous target architectures a given pass may only apply to a subset of the target system. 

\begin{figure}[!t]
\centering
\includegraphics[width=60mm]{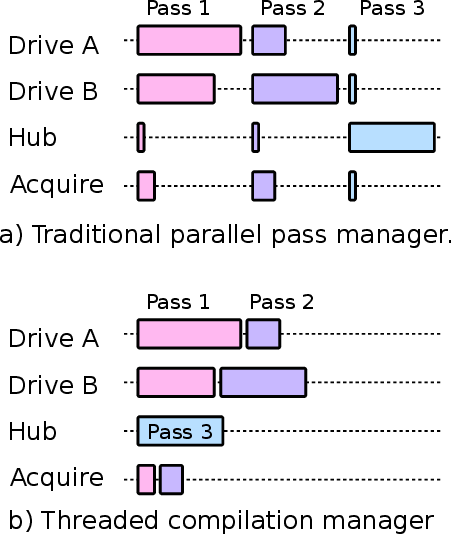}
\caption{Example pass timing for two styles of pass manager. The top figure a) shows a traditional parallel pass manager operating on the style of multiple instrument module input created by the \textit{qe-compiler}, while the bottom figure b) shows the timing using our threaded compilation manager.}
\label{fig:threaded-compilation}
\end{figure}

To solve this problem we have introduced a threaded compilation manager. The threaded compilation manager enables each target instrument to be provided with the module for that instrument and to apply an independent pass manager to that module in parallel with all other instruments. This means that passes that we know will never apply to modules for a particular instrument type need never be run on those modules. Instead, passes appropriate to that instrument can be run in parallel. Each \texttt{Target} controls its lowering pipeline and applies it only to its corresponding module. Figure \ref{fig:threaded-compilation} a) shows a standard threaded pass manager being applied to a heterogeneous target, while b) shows a threaded compilation manager being applied to the same system and input.

\section{Payload} \label{sec:payload}

The compiler produces a default payload called a quantum executable module (file extension \textit{qem}) that is a zip file containing all of the payload artifacts added by targets during compilation in addition to a standard manifest. The payload class contains an unordered map of filenames to file contents stored as strings. This makes it very simple for targets to add output data to the payload, including debug information. Adding data to the payload in a multi-threaded fashion is supported and is protected using mutex locks. The Payload class is extensible, and custom Payloads can be created using inheritance and registered with the compiler at build time.

\section{Python Interfaces} \label{sec:python-interface}

The compiler contains two types of Python interfaces. The first is the front-end interface that enables calling the compiler on an OpenQASM 3 input string and obtaining a compiled payload as a string of bytes. An additional interface is included to support parameter binding and linking. Pre-compiled payloads can be \textit{linked} with a set of parameter values, enabling payloads to be updated with new values without requiring full recompilation.

The second Python interface exposed by the compiler is MLIR-level Python bindings for all dialects supported by the compiler. These allow the creation of in-memory MLIR objects, and parsing from and printing to strings and files containing these specialized quantum engine dialects. Our goal with these is to enable lower-level access to the compiler to build calibration routines and pipelines and to enable user-level pulse access to the compiler. These are easily extensible for any new operations or dialects added to the compiler.

Compilers (including the LLVM framework the \texttt{qe-compiler} is based upon) are typically not designed to be called multiple times without reloading, which does not interact well with the calling conventions of Python. To get around this the compiler is instantiated inside its own Python subprocess. This complicates debugging to some extent but enables repeatedly calling the compiler from a single-parent Python process.

\section{Diagnostics} \label{sec:diagnostics}

The compiler also supports emitting diagnostics. These messages are categorized into various user-friendly exceptions. The Python interface (Section~\ref{sec:python-interface}) also handles these error categories and raises them to users as Python exceptions. This propagates error diagnostics created in the core C++ compiler code and makes them accessible to users. We continue to work towards creating better diagnostics and user errors to make the compiler more user-friendly.

All \texttt{Target} objects have interfaces for adding and getting diagnostics and contain a list of diagnostics that have been created. This enables any target pass or other structure with access to a target to easily add diagnostics. The presence of target diagnostics is then checked after the target pipelines are run to forward them to users using the diagnostic callback provided by the Python interface (Section \ref{sec:python-interface}).

\section{Conclusion} \label{sec:conclusion}

In this paper, we have described the design and architecture of the \textit{qe-compiler}, an open-source backend quantum control system compiler. The \textit{qe-compiler} is built on the LLVM MLIR project and defines several custom IRs/dialects used for creating executable binaries for quantum control systems. It contains a flexible and extensible target system, supports input in both MLIR and OpenQASM 3 formats, and has a Python interface for easing integration with higher-level quantum engine software stacks.

In future, we plan to continue to improve the \textit{qe-compiler}'s performance on large-scale circuits and systems by extracting more parallelism and reducing data copying within the compiler. We will also continue to increase support for classical computation. Finally, we plan to continue focusing on improving support for higher-level workloads including Probabilistic Error Amplification (PEA), Probabilistic Error Cancellation (PEC), and Quantum Error Correction (QEC).

\section{Acknowledgements} \label{sec:ack}

The authors would like to acknowledge the many important contributions to the \textit{qe-compiler} project from others within the IBM Quantum organization, not limited to, but including: Blake Johnson, Kevin Hartman, Ali Javadi-Abhari, Kevin Krsulich, Hoss Ajallooiean, Steven Casagrande, Javier G. Sogo, Kevin J. Sung, Jonathan Wildstrom, Oliver Dial, and Andrew Wack, as well as the contributions of former team members: Lauren Capelluto, Zachary Schoenfeld, and Stefan Teleman.

\printbibliography 

\end{document}